\documentclass[sigconf]{acmart}

\usepackage{svg}
\usepackage{multirow}
\usepackage{amsfonts}
\AtBeginDocument{%
  \providecommand\BibTeX{{%
    \normalfont B\kern-0.5em{\scshape i\kern-0.25em b}\kern-0.8em\TeX}}}

\setcopyright{acmlicensed}
\copyrightyear{2024}
\acmYear{2024}
\acmDOI{XXXXXXX.XXXXXXX}

\acmConference[ACM KDD '24 Workshops]{Make sure to enter the correct
  conference title from your rights confirmation emai}{August 25--29, 2024}{Barcellona, Spain}
\acmBooktitle{ACM KDD 2024 Workshops, Workshop on Uncertainty Reasoning and Quantification in Decision Making, August 25--29, 2024, Barcellona, Spain} 
\acmISBN{978-1-4503-XXXX-X/18/06}

\begin{document}

\title{Uncertainty-aware segmentation for rainfall prediction post processing}

\author{Simone Monaco}
\email{simone.monaco@polito.it}
\affiliation{%
  \institution{Department of Control and Computer Engineering, Politecnico di Torino}
  \streetaddress{Corso Castelfidardo, 39}
  \city{Torino}
  \country{Italy}
  \postcode{10129}
}
\orcid{0000-0003-4948-6120}

\author{Luca Monaco}
\email{luca.monaco@polito.it}
\affiliation{%
  \institution{Department of Environmental Engineering, Politecnico di Torino}
  \streetaddress{Corso Castelfidardo, 39}
  \city{Torino}
  \country{Italy}
  \postcode{10129}
}

\author{Daniele Apiletti}
\email{daniele.apiletti@polito.it}
\affiliation{%
  \institution{Department of Control and Computer Engineering, Politecnico di Torino}
  \streetaddress{Corso Castelfidardo, 39}
  \city{Torino}
  \country{Italy}
  \postcode{10129}
}

\renewcommand{\shortauthors}{Monaco et al.}

\begin{abstract}
    Accurate precipitation forecasts are crucial for applications such as flood management, agricultural planning, water resource allocation, and weather warnings. Despite advances in numerical weather prediction (NWP) models, they still exhibit significant biases and uncertainties, especially at high spatial and temporal resolutions. To address these limitations, we explore uncertainty-aware deep learning models for post-processing daily cumulative quantitative precipitation forecasts to obtain forecast uncertainties that lead to a better trade-off between accuracy and reliability. Our study compares different state-of-the-art models, and we propose a variant of the well-known SDE-Net, called SDE U-Net, tailored to segmentation problems like ours. We evaluate its performance for both typical and intense precipitation events.

    Our results show that all deep learning models significantly outperform the average baseline NWP solution, with our implementation of the SDE U-Net showing the best trade-off between accuracy and reliability. Integrating these models, which account for uncertainty, into operational forecasting systems can improve decision-making and preparedness for weather-related events.
\end{abstract}






\maketitle

\section{Introduction}

Accurate precipitation forecasts are essential for flood management, agricultural planning, water resource allocation, and weather warnings. Despite significant advancements in numerical weather prediction (NWP) models, these models still exhibit biases and uncertainties, especially at high spatial and temporal resolutions. This is due to the complex, nonlinear nature of atmospheric processes and inherent approximations in NWP models~\cite{bauer2015quiet,rasp2020weatherbench}. The direct model output (DMO) of NWPs is highly sensitive to initial conditions, boundary conditions, and parameterization schemes (e.g., orography). Consequently, predictions are incomplete without a characterization of the associated uncertainty~\cite{demargne2014science,TheScienceofNOAAsOperationalHydrologicEnsembleForecastService}. For instance, forecast uncertainty is crucial for the Italian Civil Protection in issuing localized severe weather warnings.

Post-processing techniques have been developed to mitigate NWP limitations and improve prediction reliability. Traditional statistical methods like model output statistics (MOS) and ensemble model output statistics (EMOS) have been somewhat successful, but often fail to capture the complexity of precipitation patterns and uncertainties~\cite{gneiting2007probabilistic,scheuerer2015variogram}.

Recently, machine learning (ML) has shown remarkable results in improving weather forecasts by processing large datasets and recognizing complex patterns that conventional methods struggle with~\cite{schultz2021can,vandal2018deepsd}.

Our contributions focus on enhancing the reliability and consistency of rainfall forecast uncertainty estimates through post-processing daily cumulative quantitative precipitation forecasts (QPF) from NWPs in northwestern Italy. 
We aim to improve prediction accuracy while ensuring reliable uncertainty estimates in precipitation forecasts. By reinterpreting rainfall estimation as an image segmentation task, we explore the application of various deep-learning approaches to develop a post-processing tool that integrates forecasts from multiple NWP models. Alongside state-of-the-art solutions, we introduced SDE U-Net, a variant of SDE-Net.~\cite{kong2020sde}, specifically designed for segmentation tasks.

This multi-model approach leverages the strengths of individual numerical models, combining them to enhance overall forecast accuracy and reliability~\cite{gagne2014machine}. We then comprehensively evaluated the proposed algorithms, particularly focusing on uncertainty estimation for typical and intense weather events. Our analysis addresses the accuracy-reliability tradeoff, balancing confidence in model predictions with the risk of forecasts missing the physical outcomes.

The post-processing systems investigated in this work can be integrated into operational forecasting systems, leading to more informed decision-making and better preparation for weather-related events.

\section{Background}
\label{sec:background}
Uncertainty can have different sources. In a machine learning context, the definitions of aleatoric and epistemic uncertainties help us understand and manage the limitations and reliability of our models' predictions. Aleatoric uncertainty is due to inherent noise in the data: this type of uncertainty is present in the observations and cannot be reduced even if we collect more data. It arises from the natural variability in the data generation process. Epistemic uncertainty reflects the model's uncertainty about its predictions due to insufficient training data or limited model capacity \cite{kendall2017uncertainties}.

\subsection{Related Works}
One of the most popular research directions for quantifying uncertainty in neural networks involves Bayesian neural networks (BNNs) ~\cite{denker1990transforming,mackay1992practical}, which quantify prediction uncertainty by imposing probability distributions over model parameters instead of using point estimates. While BNNs provide a principled method for quantifying uncertainty, the exact derivation of parameter posteriors is often computationally difficult, especially for large input data sets, such as in computer vision tasks.

Among the non-Bayesian approaches, a prominent method in this category is model ensembling~\cite{lakshminarayanan2017simple}, in which multiple deep neural networks (DNNs) with different initialization are trained and statistics on their predictions are generated for uncertainty estimation. However, training an ensemble of DNNs can be prohibitively expensive.

Other non-Bayesian methods~\cite{geifman2018bias} often mix aleatory uncertainty with epistemic uncertainty. Separating these two sources of uncertainty is crucial for many tasks~\cite{abdar2021review}. SDE-Net~\cite{kong2020sde} addresses this problem by introducing a Brownian motion term into the network architecture to capture epistemic uncertainty and view DNN transformations as state evolution in a stochastic dynamical system. However, this architecture is demonstrated on simple classification and regression tasks with tabular data and cannot be directly applied to segmentation tasks and rainfall prediction without modifications.

Several studies have used Monte Carlo (MC) dropout to estimate uncertainty. Wang et al.~\cite{wang2019aleatoric} analyzed the epistemic and aleatory uncertainty for CNN-based medical image segmentation at both pixel and structural levels.

To our knowledge, only a few works have provided estimates of uncertainties in QPF post-processing. Moosavi et al.~\cite{moosavi2021machine} have recently applied machine learning strategies to estimate and predict NWP errors in precipitation forecasting. Unfortunately, it is specific to the Weather Research and Forecasting (WRF) model and may not generalize well to other weather models.

\section{Methods}
\label{sec:method}
We can formulate our task below with a double interpretation. In the \textit{deterministic interpretation}, given a true precipitation map $P$ for a given event and a set of $n$ \textit{imperfect} predictions $\{P_i\}_{i=1,\ldots n}$ which are results of as many different NWP models, our deep learning algorithm — represented as a parametric function of weights $\theta$ — must produce an output $\hat P$ of the form
\begin{equation}
\hat{P}=f(\{P_i\};\theta) 
\label{eq:model_definition}
\end{equation}
such that the distance function
\begin{equation}
    \mathcal{L} = ||P - \hat P||_2
\end{equation}
is minimized. This equation presents an $L_2$ loss function, but other alternatives can be employed as needed.

Alternatively, from a \textit{probabilistic} point of view, we can think of the NWP outcomes $P_i$ as different i.i.d. samples from a distribution of a stochastic process of the form
\begin{equation}
    P_i = P + \delta p_i
\end{equation}
Where the $\delta p_i$ represents the epistemic error provided by each numerical model. In this framework, we expect an uncertainty in the model prediction $\hat P$ due to the type of input it was trained on. At the same time, we expect some aleatoric uncertainty due to inherent measure errors in observational data. In this work, we will not directly distinguish between the two and will just provide overall forecast uncertainty estimates where we consider both sources of error.

Conventional deep learning models are generally used deterministically, providing no access to prediction uncertainty. To address this limitation, we propose to reformulate the problem by replacing the parametric model $f$ with a variant that can produce a distribution of outcomes instead of a single value. In other words, the model prediction can be represented as a sample from this distribution:
\begin{equation}
    \hat{P} \sim \tilde f(\{P_i\};\theta) 
\end{equation}
Where $\tilde f$ represents the variational model. Given $\bar Y = \{\hat P_i\}_n$ a set of $n$ samples from the predictive distribution, we can define the prediction intervals (PIs) with a confidence level of $\gamma \in [0, 1)$ as the range $[l(\bar Y), u(\bar Y)]$ such that the probability $\mathcal{P}\left(l(\bar Y) < \hat P_{n+1} < u(\bar Y)\right) = \gamma$, which indicate the expected error between the prediction and the actual targets. A large PI indicates greater uncertainty in the model’s predictions. While the actual precipitation value is likely to be within the specified interval, the predictions may not be very accurate. Essentially, a large PI indicates that the model has less confidence in its predictions, reflecting greater variability in the input data or inherent challenges in the prediction process.

Conversely, a small PI indicates a higher confidence in the model's predictions, which suggests that the actual precipitation value is likely to be very close to the predicted value. However, this also increases the risk that the actual values will lie outside these PIs. The optimal PI range, therefore, depends heavily on the practical application and is a tradeoff between accuracy and reliability.

In the context of rainfall prediction, NWP simulations $P_i$ typically exhibit a large PI due to varying mathematical assumptions in the different models. While this broad PI is beneficial for capturing intense meteorological events, it can also lead to excessive uncertainty. Ideally, a refined model should reduce this range while maintaining sufficient width to capture significant weather events effectively.

\subsection{Case study}

Our study aims to estimate forecast uncertainties in daily cumulative QPF over northwestern Italy, specifically focusing on the Piemonte and Valle d'Aosta regions over 24 hours. These areas present a particular challenge for precipitation forecasting due to their varied topography, significantly influencing local precipitation patterns.

To address this task, we compiled a dataset of gridded daily cumulative precipitation observations from ground stations provided by ARPA Piemonte, covering the Area of Interest (AoI) with a spatial resolution of approximately 12 km. These observations, namely the $P_i$s, are interpreted as images of size $L \times W$, with each pixel being the precipitation within the associated land area. For each real observation, $n$ NWP outcomes are collected and gridded to match the shape of the ground truth, producing an image of size $L \times W \times n$ when stacked together along the channel axis.

\subsection{Deep learning architectures}

Within this framework, the problem can be phrased as a segmentation task. We chose a U-Net architecture~\cite{ronneberger2015u} as our deterministic backbone network. Despite the availability of many newer alternatives, U-Net remains extremely popular in fields such as medical imaging, remote sensor analysis, and diffusion models~\cite{9446143,peebles2023scalable}.
Its skip-connected encoder-decoder structure is particularly well suited for capturing both local and global contexts, making it ideal for our task, for which we have experimentally found that other more complex architectures do not yield remarkable results.
However, it is worth noting that our choice of U-Net is not crucial for the subsequent analysis. All the model changes we will present can also be applied to other segmentation backbones.

Based on this, we have developed several models for segmentation under uncertainty that incorporate the best-known strategies from the literature to achieve this property for different tasks. In the following sections, we briefly introduce these models and highlight our contributions to the development of some of them.

\subsubsection{Monte Carlo Dropout U-Net}\label{model:mcd}
Henceforth MCD U-Net, this approach enhances the backbone model with a Monte Carlo Dropout (MCD) strategy~\cite{gal2016dropout}. Dropout, originally introduced as a regularization procedure, involves randomly discarding a subset of neurons during training to prevent overfitting and improve the generalization of the model. In MCD, this concept is extended to the testing phase to estimate uncertainty, as the dropout at the time of inference can be considered as a Bayesian approximation. Dropout layers are applied during inference, and multiple forward passes are performed to generate a distribution of predictions. The variance of these predictions is then a measure of the uncertainty of the model.

\subsubsection{Deep Ensemble U-Net}\label{model:Ens}
Henceforth Ens U-Net, in this technique, several U-Net models are trained independently of each other with different initializations~\cite{lakshminarayanan2017simple}. As with the previous method, this approach also leads to variability in the model results, although the number of trained models limits the possible different output patterns.

\subsubsection{SDE U-Net}\label{model:SDE}
SDE-Net was recently proposed by Kong et al.~\cite{kong2020sde} to integrate Stochastic Differential Equations (SDEs) into deep learning models for capturing uncertainty. Neural networks can be viewed as continuous-time transformations of input dynamics, with model epistemic uncertainty accessed by viewing this process as a stochastic dynamical system governed by the following stochastic differential equation:
\begin{equation}\label{eq:sde_equation}
    dx_t = f(x_t, t; \theta_f)dt + g(x_0; \theta_g)dW_t
\end{equation}
Here, the diffusion term $g$ modulates the Brownian motion $dW_t$, representing the stochastic component of the process. The parametric functions $f(\cdot; \theta_f)$ and $g(\cdot; \theta_g)$ are two neural networks trained to model aleatoric and epistemic uncertainty, respectively. The training strategy ensures that $g$ provides a small variance for data within the training distribution and a large variance for data outside it. This is obtained by addressing the following objective function:
\begin{equation}
    \min_{\theta_f} \mathbb{E}\left[\mathcal{L}(x_T)\right] + \min_{\theta_g} \mathbb{E}_{x_0} \left[g(x_0; \theta_g)\right] + \max_{\theta_g} \mathbb{E}_{\tilde x_0} \left[g(\tilde x_0; \theta_g)\right],
\end{equation}
where $\mathcal{L}$ is the task-dependent loss function enforcing stochastic process' terminal outcome $x_T$ to approach the target prediction and $\tilde x0$ is an out-of-distribution sample obtained by adding Gaussian noise to the initial state $x_0$ sampled from the training data.

The original implementation of SDE-Net develops the input-output system over the time interval $[0, T]$ using an Euler-Maruyama scheme, where the two components of the equation~\ref{eq:sde_equation} are added iteratively with a fixed step size. This allows using the same networks at each time step, reducing the overall number of weights.

Extending this strategy to the U-Net architecture is a challenge because the main advantage of U-Net lies in its networked encoder-decoder structure. In U-Net, the input signal going into each encoder block generates an output that serves as the input for the next encoder block and is also passed to the corresponding decoder block via skip connections. These blocks have different input and output channels, which makes it impossible to share weights between them. To incorporate the SDE-Net strategy, we set the number of time splits to match the number of encoder blocks. For each encoder block, we construct a diffusion block placed at each skip connection and simulate an integration step at each encoder-decoder exchange. 

This network is trained using the strategy proposed in~\cite{kong2020sde} to assign higher uncertainty to out-of-distribution inputs, enabling effective uncertainty quantification in segmentation tasks while preserving the essential U-Net structure.
\subsection{Experimental design and Validation metrics}

To measure the benefits of uncertainty-aware architectures in rainfall prediction tasks, we train the models to reconstruct precipitation maps from different \textit{typical} events. In contrast, we also collect a set of events labelled as \textit{intensive} and separated from the training data. Intensive events are all those where the maximum recorded precipitation within the RoI (Region of Interest) exceeds the 99th percentile for the corresponding season. Further insights are given in the following section. Based on this separation, we expect a deep learning model to perform better when evaluated on typical events, while the performance degrades for intense events. 
We compare the uncertainty provided by a PoorMan's Ensemble (average of NWP forecasts), which is our benchmark, with the forecast uncertainty from each considered machine learning model. To get a basic uncertainty estimate, we use normalized rMSE, while to quantify the trade-off between accuracy and reliability, we introduce a coverage-length-based criterion (CLC) as defined in ~\cite{khosravi2010construction}
\begin{equation}
    CLC = \frac{NMPIL}{\sigma\left(PICP,\eta,\mu\right)},
\end{equation}
where $\sigma$ is a sigmoid function with scaling parameter $\eta$ and translation parameter $\mu$:
\begin{equation}
    \sigma\left(PICP,\eta,\mu\right) = \frac{1}{1+e^{-\eta(PICP - \mu)}}
\end{equation}
We aim to achieve low values of Normalized Mean Prediction Interval Length (NMPIL), as it indicates a narrower spread in ensemble predictions, which we seek to minimize for more meaningful and useful predictions. However, reducing NMPIL negatively affects the coverage of Prediction Intervals (PIs), resulting in an undesirable number of predictions falling outside the PIs. To address this issue, we aim for high PI Coverage Probability (PICP) values, which measure the proportion of target values within the prediction interval. Consequently, we strive for the smallest possible values of CLC. The parameter $\eta$ controls the penalty when PICP falls below the minimum acceptable level $\mu$.

Ideally, the threshold value of acceptability $\mu$ should be as close as possible to 1, so we set it to a reasonably high value, namely $\mu=\gamma=0.95$. In \cite{bishop1995neural} the parameter $\eta$ is explored in the context of neural networks training in order to study learning sensitivity and dynamics, leading to a useful range $1<\eta<10$: this contribution can be applied also in other contexts such as CLC, so we examine the behaviour of CLC as a function of $\eta$ ranging from 0 to 12, which slightly extends the suggested range. Of course, predictions falling outside a PI with $\mu=0.95$ should be strongly penalized, so we are particularly interested in the CLC values for high $\eta$ (i.e. around 10). Accordingly, we provide tabular values for rMSE, PICP, NMPIL, and CLC with $\eta=12$, the maximum value considered in our analysis.

MCD U-Net and SDE U-Net use 20 sampled predictions, while Ens U-Net (\ref{model:Ens}) is based on 5 repetitions, i.e., as many ensemble models. This approach estimates forecast uncertainty for each model, reflecting epistemic error. We then repeat the process in a 9-fold cross-validation to ensure statistical significance, accessing aleatoric uncertainty. This involves training the models on nine different training-validation-test subsets, derived using weather physics considerations detailed in subsection \ref{sec:datasetbuild}. Uncertainty estimates for each validation metric $VM$ are provided as:
\begin{equation}
    VM = {VM}_{repetitions} \pm {VM}_{9-fold\ CV}.
\end{equation}
\section{Experiments}
\label{sec:experiments}
\subsection{Dataset building}\label{sec:datasetbuild}

As previously mentioned, the dataset includes observations from ground stations provided by ARPA Piemonte, preprocessed using optimum interpolation~\cite{gandin1963objective} to generate images on a fixed grid. These observations span from 1957 to the present, providing a continuous and comprehensive record of precipitation across various meteorological conditions. The dataset also includes precipitation forecasts from four NWP models: BOLAM-CNR~\cite{buzzi1994validation}, ECMWF-IFS~\cite{ecmwf2016ifs}, COSMO-2I~\cite{baldauf2011operational}, and COSMO-5M~\cite{doms2018description}.

Events were classified as ``intense" if their spatial maximum precipitation exceeded the seasonal 99th percentile, with thresholds of 64.58mm in winter, 95.71mm in spring, 93.26mm in summer, and 140.40mm in autumn. This classification resulted in 436 events, 40 of which were marked as intense and set aside during the training phase.

For typical events, we applied K-means clustering based on the variability-average plane to categorize events into convective, stratiform, and intermediate types. These types are characterized by high spatial variability and low spatial average precipitation, low spatial variability and high spatial average precipitation, and intermediate characteristics, respectively~\cite{houze1997stratiform,sorooshian2002evaluation}. The dataset was then divided into nine distinct training-validation-test subsets, ensuring uniform event type distribution across all subsets.
The code for our experiments is available online~\footnote{https://github.com/simone7monaco/rainfall-prediction}, while the dataset can be shared upon request.

\subsection{Results}
\begin{figure}
    \centering
    \includegraphics[width=\linewidth]{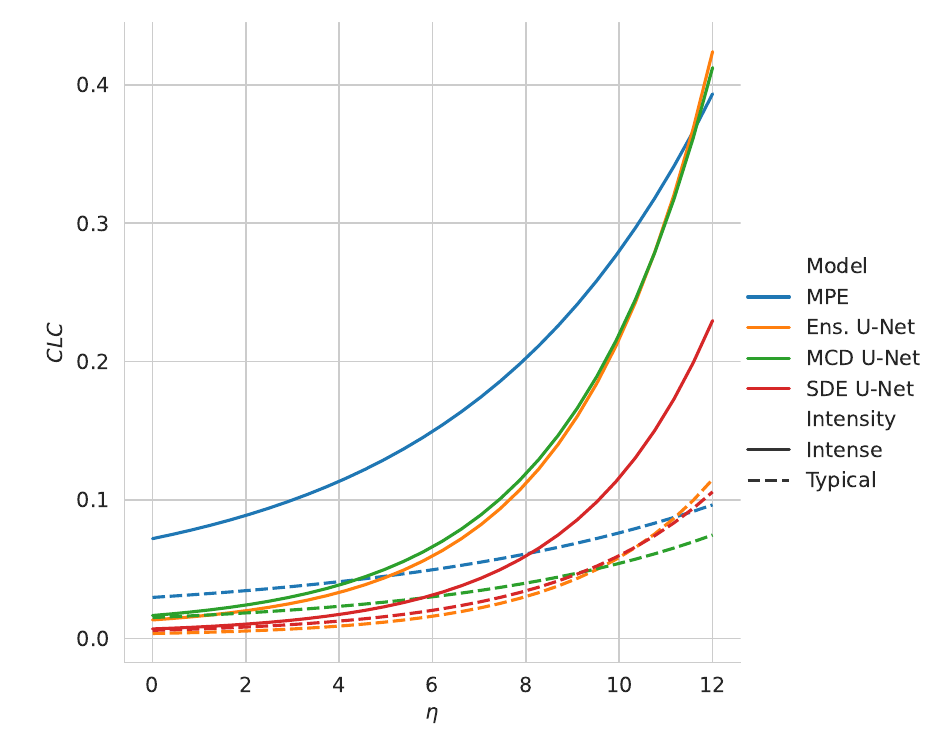}
    \Description[CLC Score of the model over the parameter $\eta$]{}
    \caption{CLC Score of the model over the parameter $\eta$}
    \label{fig:clc}
\end{figure}

\begin{table*}
    \centering
    \resizebox{\linewidth}{!}{%
        \begin{tabular}{llllllllll}
        \toprule
         & \# Parameters (M) & \multicolumn{2}{c}{rMSE ($\times 100$) $\downarrow$} & \multicolumn{2}{c}{PICP $\uparrow$} & \multicolumn{2}{c}{NMPIL ($\times 100$) $\downarrow$} & \multicolumn{2}{c}{CLC ($\mu=0.95$, $\eta=12$) $\downarrow$} \\
         &  & Typical & Intense & Typical & Intense & Typical & Intense & Typical & Intense \\
        \midrule
        PME & - & $1.156$ & $3.152$ & $\mathbf{0.808}$ & $\mathbf{0.759}$ & $1.483$ & $3.605$ & $0.096$ & $0.393$ \\
        Ens. U-Net & $5 \times 31.0$ & $\mathbf{0.779 {\scriptstyle \pm 0.113 }}$ & $2.746 {\scriptstyle \pm 0.037 }$ & $0.626 {\scriptstyle \pm 0.032 }$ & $0.608 {\scriptstyle \pm 0.005 }$ & $\mathbf{0.222 {\scriptstyle \pm 0.074 }}$ & $0.681 {\scriptstyle \pm 0.053 }$ & $0.110 {\scriptstyle \pm 0.036 }$ & $0.419 {\scriptstyle \pm 0.020 }$ \\
        MCD U-Net & $31.0$ & $0.834 {\scriptstyle \pm 0.091 }$ & $2.849 {\scriptstyle \pm 0.063 }$ & $0.787 {\scriptstyle \pm 0.027 }$ & $0.628 {\scriptstyle \pm 0.004 }$ & $0.791 {\scriptstyle \pm 0.074 }$ & $0.804 {\scriptstyle \pm 0.048 }$ & $\mathbf{0.065 {\scriptstyle \pm 0.017 }}$ & $0.393 {\scriptstyle \pm 0.026 }$ \\
        SDE U-Net & $12.4$ & $0.815 {\scriptstyle \pm 0.167 }$ & $\mathbf{2.637 {\scriptstyle \pm 0.016 }}$ & $0.665 {\scriptstyle \pm 0.016 }$ & $0.602 {\scriptstyle \pm 0.001 }$ & $0.299 {\scriptstyle \pm 0.011 }$ & $\mathbf{0.345 {\scriptstyle \pm 0.019 }}$ & $0.095 {\scriptstyle \pm 0.014 }$ & $\mathbf{0.229 {\scriptstyle \pm 0.009} }$ \\
    
        \bottomrule
        \end{tabular}
    }%

    \caption{rMSE, PICP, NMPIL e CLC at $\eta=12$ in deep learning models vs PoorMan's Ensemble. Up and down arrows indicate whether the best value is the higher or the lower.}
    \label{tab:agg_results}
\end{table*}
Figure \ref{fig:clc} displays the behaviour of CLC for $\mu=0.95$ and $\eta$ values ranging from 0 to 12 for typical and intense events, comparing all deep learning models to the average of the weather models (PoorMan's Ensemble, PME). Error bars are omitted for readability.
Smaller CLC values indicate a better trade-off between accuracy and reliability, achieved through low NMPIL and high PCIP values, particularly at higher $\eta$ values. Table~\ref{tab:agg_results} summarizes the results of our analysis in terms of CLC with $\mu=0.95$ and $\eta=12$, alongside rMSE, PICP, and NMPIL for all uncertainty-aware models compared to PME.

As expected, rMSE is higher for intense events than for typical events, with all deep learning models significantly outperforming PME. The low uncertainty in rMSE estimates highlights the model with the best rMSE. Ens U-Net achieves the lowest rMSE for typical events ($8.15\cdot 10^{-3}$), while SDE U-Net achieves the lowest for intense events ($2.637\cdot 10^{-2}$), indicating higher prediction accuracy.

The PICP column shows that PME has better percentage coverage than the learning models by 10 to 15\%, but at the cost of much wider prediction intervals, as indicated by the NMPIL column. PME has NMPIL values of $1.489\cdot 10^{-2}$ for typical events and $3.605\cdot 10^{-2}$ for intense events, compared to the deep learning models' range of $2\cdot 10^{-3}$ to $8\cdot 10^{-3}$ for both event types. This suggests that PME predictions are reliable but lack accuracy.

However, the trade-off between accuracy and reliability differs for typical and intense events when considering CLC at $\eta=12$. For typical events, CLC indicates no substantial advantage for deep learning models over PME ($0.09<{CLC}<0.11$ for PME, Ens U-Net, and SDE U-Net), although MCD U-Net shows a slight improvement (0.065). For intense events, SDE U-Net has the lowest CLC value (0.229), with the differences between PME, MCD U-Net, and Ens U-Net being negligible. While SDE-UNet does not have the highest PICP, its primary advantage is the small prediction intervals, as reflected in NMPIL, especially for intense events ($3.45\cdot 10^{-3}$). This results in a highly favourable accuracy-reliability trade-off, as shown by CLC.

For $8 < \eta < 12$, PME consistently shows higher CLC values for typical events compared to deep learning models, though the difference is minor. SDE U-Net and Ens U-Net perform comparably or slightly better than MCD U-Net around $\eta = 9$. For intense events, PME consistently underperforms against the deep learning models. MCD U-Net and Ens U-Net have similar CLC values, but SDE U-Net demonstrates the best accuracy-reliability tradeoff.



Overall, these results highlight the effectiveness of deep learning models, particularly the SDE U-Net, in providing accurate and precise rainfall predictions while maintaining a reasonable level of uncertainty quantification.

\section{Conclusions}
\label{sec:Conclusions}
\balance
To summarise, our study demonstrates the effectiveness of deep learning solutions to improve the accuracy and reliability of NWP post-processing systems for precipitation forecasts. By evaluating both typical and intense precipitation events, we found that all deep learning models significantly outperformed the average baseline NWP solution, with our implementation of SDE-UNet showing the best trade-off between accuracy and reliability.

Integrating these models, which account for uncertainty, into operational forecasting systems can improve decision-making and better preparation for weather-related events. Future work will focus on refining these models and exploring alternatives to achieve more comprehensive results in predicting precipitation while accounting for uncertainty.

\begin{acks}
This work is part of the project NODES, funded by the Italian MUR (Ministry of University and Research) under M4C2 1.5 of the PNRR (National Plan for Recovery and Resilience) with grant agreement no. ECS00000036.
The SmartData@PoliTO research centre of Politecnico di Torino, Italy has partially funded this work.
\end{acks}

\bibliographystyle{ACM-Reference-Format}
\bibliography{'biblio}

\end{document}